\documentclass[11pt]{article}
\usepackage{graphics}
\usepackage{epsfig}

\def\be{\begin{equation}}  
\def\bea{\begin{eqnarray}}  
\def\ee{\end{equation}}     
\def\eea{\end{eqnarray}}     
\def\gsim{\mathrel{\raise.3ex\hbox{$>$\kern-.75em\lower1ex\hbox{$\sim$}}}}
\def\lsim{\mathrel{\raise.3ex\hbox{$<$\kern-.75em\lower1ex\hbox{$\sim$}}}}

\begin{document}

\vskip .7cm

\vspace{30cm}

\begin{center}

{\Large \bf Evolutionary and structural properties of \\
\vskip .2cm
                  mirror star MACHOs}

\vskip .7cm

{\large Zurab Berezhiani $^{a,}$\footnote{E-mail:
     {\tt berezhiani@aquila.infn.it}}}, ~ 
{\large Paolo Ciarcelluti $^{a,}$\footnote{E-mail:
     {\tt ciarcelluti@lngs.infn.it}}}, ~ 
{\large Santi Cassisi $^{b,}$\footnote{E-mail:
     {\tt cassisi@te.astro.it}}}, ~ \\
{\large Adriano Pietrinferni $^{b,}$\footnote{E-mail:
     {\tt adriano@te.astro.it}}}

\vskip .7cm

{\it 
$^a$  Dipartimento di Fisica, Universit\`a di L'Aquila, 67010 Coppito  
AQ, and  \\
  INFN, Laboratori Nazionali del Gran Sasso, 67010 Assergi AQ, 
Italy \\ 
$^b$  INAF, Osservatorio Astronomico di Collurania, 64100 Teramo, Italy
}

\end{center}

\setcounter{footnote}{0}
\setcounter{page}{1}

\vskip .3cm

\begin{abstract}
There can exist a hidden sector of the Universe in the form of parallel 
``mirror'' world which has the same particle physics as the observable 
world and interacts with the latter only gravitationally. 
Big Bang Nucleosynthesis bounds demand that the mirror sector should 
have a smaller temperature than the ordinary one. 
This implies that the mirror matter could play a role of dark matter, and in 
addition its chemical content should be dominated by helium. 
Here we study the evolutionary and structural properties of the mirror stars 
which essentially are similar to that of the ordinary stars but with higher 
helium contents. 
Being  invisible in terms of photons, they could be observed only as 
MACHOs in the microlensing experiments. 
Using a numerical code, we compute evolution of stars with large helium 
abundances ($Y = 0.30-0.80$)
and a wide range of masses, from 0.5 to 10 $ M_\odot $. 
We found that helium dominated mirror star should have much faster 
evolutionary time (up to a factor $\sim$ 30) than the ordinary star with the 
same mass. 
In addition, we show the diagrams of luminosities, effective temperatures, 
central temperatures and densities, and compute the masses of the He 
core at ignition and the minimum mass for carbon ignition, for different 
chemical compositions. 
The general conclusion is that mirror stars evolve faster as compared to 
ordinary ones, and explode earlier as type II supernovae, thus enriching 
the galactic halo of processed mirror gas with higher metallicity, 
with implications for MACHO observations and galaxy evolution. 
\end{abstract}

\vspace{0.3cm}

%---------------------------------------------------------------------------------

\section{Introduction}

At present one of the possible candidates of dark matter is mirror matter 
(for reviews, see refs.~\cite{books,IJMPA-F,IJMPA-B}). 
The idea of mirror world --  a parallel hidden sector of particles which is an 
exact duplicate of the observable particle sector -- has been suggested a 
long time ago by Lee and Yang \cite{mirror}, and various phenomenological 
aspects of this hypothesis were first discussed by Kobzarev, Okun and 
Pomeranchuk \cite{KOP}. 
Subsequently, different implications of mirror world for the particle physics, 
astrophysics and cosmology were addressed in a series of papers 
\cite{pavsic}-\cite{dundee}. 

The basic concept is to have a hidden mirror (M) sector of the Universe 
which has exactly the same particle physics as that of  the ordinary (O) 
sector. 
In other words, this theory is given by the product  $G \times G'$ 
of two identical gauge factors, which could naturally emerge e.g. in the 
context of $E_8 \times E'_8$ superstring. 
(From now on all particles and parameters of the mirror sector will be marked 
by $'$ to distinguish from the ones belonging to the observable or ordinary 
world.)
A mirror parity $G \leftrightarrow G'$ under interchanging of all fields 
in corresponding representations between $G$ and $G'$ factors 
implies that both particle sectors are described by the same Lagrangians. 

In particular, since the physics of ordinary world is described by the 
Standard Model $SU(3)\times SU(2)\times U(1)$ with gauge fields (gluons, 
photons, $W$ and $Z$ bosons) coupled to ordinary quarks $q$ and leptons 
$l$, the physics of mirror sector should be described by the analogous 
gauge symmetry $SU(3)' \times SU(2)' \times U(1)'$ with 
the corresponding gauge bosons (mirror gluons, mirror photons, 
$W'$ and $Z'$ bosons) coupled to mirror quarks $q'$ 
and leptons $l'$. 
(The corresponding Lagrangian terms in explicit form and the field 
transformation properties under mirror parity were presented in 
ref.~\cite{FLV1}.) 
The extension towards grand unification is straightforward: one can consider 
e.g. gauge theory based on $SU(5)\times SU(5)^\prime$, 
$SO(10)\times SO(10)^\prime$, etc. 
In any case, the mirror parity implies that all coupling constants (gauge, 
Yukawa, Higgs) have the same pattern in both sectors and thus their 
microphysics is the same. 
Obviously, two sectors are connected by universal gravity, but they could 
communicate also by other means. 
For example, ordinary photons could have kinetic mixing with mirror photons 
\cite{Holdom}, ordinary (active) neutrinos could mix with mirror (sterile) 
neutrinos \cite{neutrino}, or two sectors could have a common flavour 
symmetry  \cite{flavor} or Peccei-Quinn symmetry \cite{assione}. 

The different cosmological implications of the mirror world were addressed 
in several earlier papers \cite{blinkhlo,dolgov_dnu,Macho}. 
The time evolution of the mirror Universe, starting from inflation and 
analyzing epochs of baryogenesis, nucleosynthesis, recombination and 
structure formation was consistently studied in ref.~\cite{bcv}. 
The results can be briefly summarized as follows. 

If the mirror sector exists, then the Universe along with the ordinary 
electrons, nucleons, neutrinos and photons, should also contain their mirror 
partners.  
One could naively think that due to mirror parity the ordinary and mirror 
particles should have the same cosmological abundances and hence the O 
and M sectors should have the same cosmological evolution. 
However, this would be in the immediate conflict with the Big Bang 
nucleosynthesis (BBN) bounds on  the effective number of extra light 
neutrinos, since the mirror photons, electrons and neutrinos would give a 
contribution to the Hubble expansion rate equivalent to the effective number 
of extra neutrinos $\Delta N_\nu \simeq 6.14$.

Therefore, the BBN bounds require that at the nucleosynthesis epoch the 
temperature of the mirror sector should be smaller than that of the ordinary 
one, $T' < T$.  
In this case the contribution of the mirror sector translates into 
$\Delta N_\nu \approx 6.14 x^4$ \cite{dolgov_dnu,bcv}, where $x = T'/T$. 
Thus, the BBN bound on $\Delta N_\nu$ implies the upper limit 
$x < 0.64 (\Delta N_\nu)^{1/4}$, with rather mild dependence on 
$\Delta N_\nu$. 
E.g. $\Delta N_\nu <0.5$ implies roughly $x < 0.5$ while $\Delta N_\nu <0.2$ 
implies $x < 0.4$.  
Hence, in the early Universe the mirror system should have a somewhat 
lower temperature than ordinary particles. 
This situation can be realized in a most plausible way under the following 
paradigm: 

A. At the ``Big Bang'' the two systems are born with different efficiency: 
namely, at the post-inflationary epoch the M sector is (re)heated at lower 
temperature than in the observable one, which can be naturally achieved in 
certain models \cite{dolgov_dnu,bcv,UHE}.

B. At temperatures below the reheating temperature two systems interact 
very weakly, so that they do not come into thermal equilibrium with each 
other after reheating.
This condition is automatically fulfilled  if the two worlds communicate only 
via gravity.
If there are some other effective couplings between the O and M particles, 
they have to be properly suppressed. 

C. Both systems expand adiabatically, without significant entropy 
production, keeping nearly constant the ratio of their temperatures $T'/T$.  

The parameter $x=T'/T$ plays an important role for describing the 
cosmological evolution of mirror world. 
As far as the mirror world is cooler than the ordinary one, $x < 1$, in the 
mirror world all key epochs as are baryogenesis, nucleosynthesis, 
recombination etc. proceed in somewhat different conditions than in 
ordinary world.  
Namely, in the mirror world the relevant processes go out of equilibrium 
earlier than in ordinary world, which has many far going implications. 

First, as far as baryogenesis epoch is concerned, the origin of the baryon 
asymmetry in both sectors is related to the same particle physics, as far as 
both sectors have identical Lagrangians. 
However, the out-of-equilibrium condition in the M sector is satisfied better 
than in O sector \cite{bcv}. 
Therefore, it is pretty plausible that asymmetry of M baryons is bigger than 
baryon asymmetry in ordinary sector and hence mirror baryons could 
constitute dark matter, or at least its significant fraction. 
The situation emerges in a particularly appealing way in the leptogenesis 
scenario due to entropy and lepton number leaking from the hotter O sector 
to the cooler M sector \cite{baryo-lepto}, which leads to 
$\Omega'_B/\Omega_B\geq 1$, up to an order of magnitude. 
This can explain the close relation between the visible and dark matter 
components in the Universe in a rather natural way 
\cite{IJMPA-B,baryo-lepto}. 
(For a somewhat different scenario see also refs.~\cite{BL-FV}). 

Second, in mirror sector radiation decouples from matter earlier than in 
ordinary one, at redshifts $z'_{\rm dec} \approx (1/x) z_{\rm dec}$, 
where $z_{\rm dec} \simeq 1100$ is the redshift of decoupling in ordinary 
sector \cite{bcv,ignavol-lss}. 
In particular, for $x < 0.3$ the mirror photons decouple before the 
matter-radiation equality, yet in the relativistic expansion epoch. 
As a result, for such small values of $x$, for the cosmological scales which 
still undergo the linear growth, the mirror baryons behave exactly the same 
way as conventional cold dark matter (CDM) \cite{bcv}. 
The exact computations show that for $x <0.3$  implications of the mirror 
baryons for the cosmic microwave background and the large scale structure 
of the Universe are practically indistinguishable from that of the CDM 
\cite{paolo}. 

Third, as far as the primordial nucleosynthesis epoch is concerned, it was 
shown in ref.~\cite{bcv} that the mirror helium abundance should be much 
larger than that of the ordinary helium, and for $x <0.3$ the mirror helium 
gives a dominant mass fraction of the mirror matter. 
The reason is simple. 
As far as $x<< 1$, then the impact of mirror sector on ordinary 
nucleosynthesis is insignificant: it is equivalent to 
$\Delta N_\nu \approx 6.14 x^4$, and so the BBN prediction for ordinary 
helium mass fraction $Y_4 \simeq 0.24$ is not affected significantly. 
However, the impact of the ordinary sector on the mirror nucleosynthesis is 
dramatic: it is equivalent to $\Delta N'_\nu \approx 6.14/x^4$, and so for 
$x=0.6-0.1$ the mirror helium mass fraction varies from $Y'_4 = 0.4-0.8$ 
\cite{bcv}. 

Concluding, mirror matter can be a viable candidate for dark matter of the 
Universe. 
It has the same microphysics as ordinary matter, but somewhat different 
cosmology, which makes it extremely interesting object for further 
investigations. 
Mirror baryons form the stable matter, exactly as their ordinary counterparts, 
and they should form atoms, molecules, and then even astrophysical 
objects, such as stars, planets, globular clusters, etc., as the ordinary 
matter does, however the chemical content of the mirror matter should be 
different from the ordinary one. 
In particular, it should be mainly helium dominated and in addition the 
heavier elements are also expected with bigger abundances than in ordinary 
world. 
Once the visible matter is built up by ordinary baryons, then the mirror 
baryons would constitute dark matter in a natural way.  
They interact with mirror photons, but not interact with ordinary photons.  
In other words, mirror baryons are dark for the ordinary observer, and mirror 
stars as well, and so the latter should be seen as Massive Astrophysical 
Compact Halo Objects (MACHOs) in the microlensing experiments 
\cite{dolgov_dnu,Macho}.\footnote{
The signatures of mirror matter in the search of meteoric event anomalies 
\cite{mir_meteor} and close-in extra solar planets \cite{mir_planet} were also 
discussed.} 
However,  if there exists small kinetic mixing between ordinary and mirror 
photons \cite{Holdom},  the mirror particles become sort of ``millicharged'' 
particles for the ordinary observer, which suggests a very appealing 
possibility of their detection in the experiments for the direct search of dark 
matter  \cite{IJMPA-F,mir_dama}. 
It is also extremely interesting that such a kinetic mixing can be 
independently tested in laboratory ``table-top'' experiments for searching the 
orthopositronium oscillation into its mirror counterpart \cite{ortho}. 
The mixing between the ordinary and mirror neutrinos \cite{neutrino} can be 
also tested in terms of active-sterile neutrino oscillations \cite{neutrino1}. 
In addition, it could provide a possible mechanism for the generation of ultra 
high energy neutrinos \cite{UHE} and for the gamma ray bursts as a result 
of explosion of the mirror supernovae \cite{mir_GRB}.\footnote{
Also the mirror axion could provide a plausible mechanism for the gamma 
ray bursts and supernova type II explosions \cite{assione}. }
Neutron -- mirror neutron oscillation could provide a very efficient 
mechanism for trasporting ultra-high energy protons and also explain their 
correlation with far distant sources like BL Lacs \cite{mir_neutron}.
Explosions of the mirror supernovae could provide a necessary energy 
budget for heating the gaseous part of the mirror matter in the galaxies and 
hence to prevent its collapse to a disk, in which case the mirror matter could 
form spheroidal halos in accord to observations \cite{mir_halo}. 
In addition, the efficiency of mirror supernovae explosions can be indirectly 
tested in the future detectors for the gravitational waves at the frequencies 
around 1 kHz. 
Therefore, one of the most interesting problems concerning the mirror world 
consists it the study of the formation and evolution of mirror stars. 

The aim of this paper is to study the evolutionary and structural properties of 
mirror stars.  
As far as the mirror world should be dominated by helium, and thus it 
contains less hydrogen, the evolutionary properties of the mirror stars 
should be significantly different from that of the ordinary stars. 
In particular, we calculate the evolutionary times of the mirror stars in the 
wide range of masses for different abundances of the mirror helium, and study 
under which conditions they could end up as supernovae.  
Apart of motivations which has been already discussed above, the possible 
concrete applications of our analysis for the testing of the mirror matter 
features  can be formulated as follows: 

\begin{itemize}
\item The mirror stellar evolution is one of the crucial ingredients in 
non-linear astrophysical processes involved in the cosmological structure 
formation.
In particular we need the stellar feedback in order to compute N-body 
simulations of structure formation at non-linear scales, and extend the 
matter power spectra calculated in our previous works \cite{paolo} to smaller 
scales. 
This would help us to understand the expected crucial differences between 
the mirror baryonic dark matter and CDM scenarios.
\item We need mirror stellar models in order to understand the 
circumstances under which in the galaxy the ordinary matter forms a disk 
while mirror matter  can form the spheroidal halo.
In this analysis it is crucial to have in view the two aspects played by the 
mirror star formation and evolution: 
first, M stars are collisionless and then do not 
dissipate the energy falling in the disk, but stay in the halo; and second, the 
heavy mirror stars could explode as supernovae and could provide the 
necessary  energy emission required for the mirror galaxy to balance the 
dissipation and avoid the formation of a mirror disk.
\item M stars are a natural candidate for the MACHOs observed in the galaxy.
At present the baryonic candidates seem unable to explain the MACHO 
galactic population \cite{freese}, and CDM has many problems to clump in 
objects of stellar masses, so that the mirror hypothesis is an interesting 
viable alternative. 
The study of the features of the mirror star evolution (together with the star 
formation) is one of the necessary ingredients for answering the question 
which fraction of the mirror matter can exist in the form of mirror MACHOs 
and which fraction in the form of gas or dust. 
\item This study is also necessary for understanding the rate and properties of 
mirror supernovae, which could be responsible for the gamma ray bursts 
observed in our sector \cite{mir_GRB} and provide gravitational waves 
observable by next generation detectors.
\item A possibly useful collateral effect of this study is that it is applicable 
to very late population of ordinary stars, when a large part of hydrogen has 
burned into helium and new stars form with much more helium in their 
chemical compositions.
\end {itemize}

The plan of the paper is as follows. 
In next section we introduce the mirror stars as MACHO candidates.
In sections 3 we present the mirror star models.
Section 4 analyzes in detail the evolutionary and structural properties of the 
stars, showing the luminosities, effective temperatures, central densities 
and temperatures, helium core masses at ignition and the minimum mass 
for carbon ignition, as functions of the primordial helium content and of the 
mass.
Finally, our main conclusions are summarized in section 5.

%---------------------------------------------------------------------------------

\section{Mirror stars as helium dominated stars}
\label{mirror_gal}

If mirror matter exists, then the existence of mirror stars, in a certain sense, 
is guaranteed by the existence of ordinary stars: given that two sectors have 
the same microphysics, stars necessarily form in both of them. 
Due to different initial conditions, $x=T'/T < 1$, two sectors have 
different cosmological evolution, and in particular, different chemical content. 
Thus, the details of the galaxy formation scenario depend on the exact 
composition of matter and they can be different in two sectors.  
However, one has to take into account that during the galaxy evolution 
some fraction of mirror baryons would fragment into stars. 

As far as the chemical contents are concerned, for a given value of $x$, the 
primordial helium abundances in ordinary and mirror sectors roughly can be 
given by the following formulas \cite{bcv} 
\begin{equation} \label{helium}
Y = {{ 2\exp[-t_N/\tau(1+x^{4})^{1/2}] } \over
         {1+\exp[\Delta m/T_W(1+x^{4})^{1/6} ] }} ~,
\end{equation}
and 
\begin{equation} \label{m_helium}
Y' = {{ 2\exp[-t_N/\tau(1+x^{-4})^{1/2}] } \over
         {1+\exp[\Delta m/T_W(1+x^{-4})^{1/6} ] }} ~,
\end{equation}
where $\tau = 887$ s is the neutron lifetime, $\Delta m  = 1.29$ MeV is the 
neutron-proton mass difference, $T_W \simeq 0.8$ MeV is a weak interaction 
freezing temperature in the standard cosmology, and $t_N \sim 200$ s is the 
cosmological timescale corresponding to the ``deuterium bottleneck'' 
temperature  $T_N\sim 0.07$ MeV. 
Therefore, for $x \ll 1$ the standard BBN predictions essentially are not 
affected. 
Namely, the smaller is $x$, the prediction for ordinary helium decreases and 
gets closer to the standard BBN result $Y\simeq 0.24$, while the prediction 
for the mirror helium  increases and at $x \to 0$ approaches $Y^\prime \to 1$.
In particular, for $x=0.6$ one has $Y^\prime \simeq 0.4$, while for $x=0.1$, 
$Y^\prime \simeq 0.8$ (for more precise computations see \cite{bcv}). 
Therefore, in two sectors the first stars are formed with different initial 
abundances of helium. 

The evolutionary and structural properties of stars  strongly depend on the 
initial chemical composition, which is fixed by the helium abundance and by 
the global amount of heavy elements (indicated as metallicity $Z$).
The primordial abundances of ordinary nuclei heavier than helium 
are estimated to be very small ($Z\sim10^{-10}$). 
This metallicity would be characteristic of the \lq{first}\rq\ stellar population, the so-called Population III stars.
Concerning the primordial metallicity of mirror matter, it can be be some 
orders of magnitude higher, however there are no reasons to expect that it 
will be relevant. 

Meanwhile, given the complexity of the physics of galaxy formation 
(this process in still to be well understood), we can make some general 
considerations. 
At a stage during the process of gravitational collapse of the protogalaxy, 
it fragments into hydrogen clouds with typical Jeans mass (for mirror matter, 
these gas clouds are rather the hydrogen-helium clouds).  
Clouds continue to cool and collapse until the opacity of the system 
becomes so high that the gas prefers to fragment into protostars. 
This complex phenomenon lead a part of the protogalactic gas to form the 
first stars (probably very massive, with $ M \sim 10^{2} $--$ 10^{3} M_\odot $).
A difficult question to address here is related to the star formation in mirror 
sector, also taking into account that its temperature/density conditions and 
chemical contents are much different from the ordinary ones. 
Clearly, the details of this process (Jeans mass, etc.) depend on these 
conditions, and hence should be different for ordinary and mirror matter 
components. 
The cooling rates are mainly determined by hydrogen atoms and molecules, 
while helium is much less effective. 
However, even in mirror sector, unless $x$ is extremely small, the number 
density of hydrogen remains significant. 

The pattern galaxy evolution features should drastically depend on the 
mirror star formation and evolution features. 
Stars play an important role: the fraction of baryonic gas involved in their 
formation becomes collisionless on galactic scales, and supernova 
explosions enrich the galaxy of processed collisional gas (stellar feedback). 
Too fast star formation in mirror component would extinct the mirror gas and 
thus could avoid that mirror baryons form disk galaxies as ordinary baryons 
do. 
If the mirror protogalaxy,  at certain stage of collapse, transforms into the 
collisionless system of the mirror stars, then it could maintain a typical 
elliptical structure.
Certainly, in this consideration also the galaxy merging process should be 
taken into account. 
Efficient merging of mirror disks mostly built up of stars, also would lead to 
ellipticals.\footnote{
In other words, one can speculate on the possibility that mirror baryons form 
mainly the elliptical galaxies. 
For a comparison, in the ordinary world the observed bright galaxies are 
mainly spiral while the elliptical galaxies account about $20 ~\%$ of them. 
Remarkably, the latter contains old stars, very little dust and shows no sign 
of active star formation.} 
As for ordinary matter, within the dark mirror halo it should typically show up, 
depending on conditions of the galaxy formation, as an observable elliptic or 
spiral galaxy, but some anomalous cases can also be possible, like certain 
types of irregular galaxies or even dark galaxies dominantly made out of 
mirror baryons. 
The central part of halo can nevertheless contain a large amount of ionized 
mirror gas and it is not excluded that it can have a quasi-spherical form, 
thus possibly avoiding the problem of cusp typical for the CDM halos. 
Even if mirror star formation is very efficient, the massive mirror stars in the 
dense central region could fast evolve and explode as supernovae, 
leaving behind compact objects like neutron stars or black holes, and 
reproducing the mirror gas and dust. 
It is interesting to understand whether these features could help in 
understanding the process of the formation of the central black holes, with 
masses $10^{6}-10^9\, M_\odot$, which are considered as main engines of 
the quasars and active galactic nuclei. 

The fact that dark matter made of mirror baryons has the property of 
clumping into compact bodies such as mirror stars leads to a natural 
explanation for the mysterious MACHOs. 
In the galactic halo (provided that it is the elliptical mirror galaxy) the 
mirror stars should be observed as MACHOs in gravitational microlensing  
\cite{Macho}. 
The MACHO collaboration \cite{macho542} studied the nature of halo dark 
matter by using this technique. 
This experiment has collected 5.7 years of data and provided statistically 
strong evidence for dark matter in the form of invisible star sized objects, 
which is what you would expect if there was a significant amount of mirror 
matter in our galaxy. 
Their maximum likelihood analysis implies a MACHO halo fraction of $20\%$ 
for a typical halo model with a $95\%$ confidence interval of $8\%$ to $50\%$.
Their most likely MACHO mass is between $0.15 M_{\odot}$ and 
$0.9M_{\odot}$ (depending on the halo model), with an average around 
$M\simeq 0.5 ~M_\odot$, which is difficult to explain in terms of the brown 
dwarfs with masses below the hydrogen ignition limit $M < 0.1 M_{\odot}$ 
or other baryonic objects\footnote{
An interesting MACHO candidate are the Very Low Mass (VLM) stars (see 
ref.~\cite{vlm} and references therein), namely stars with masses below 
$\sim 0.4 M_\odot$. 
They have very low luminosities and temperatures, and hence are very 
difficult to detect and possibly contribute to the galactic dark stars, but 
anyway their estimated abundances cannot account for the missing mass of 
the galaxies.}$^,$\footnote{
Indeed the rate of MACHO events as derived from microlensing data is still a controversial issue, as emphasized in ref.~\cite{micropuzzle} and references therein.
Clearly the outcome of the controversy is crucial for the mirror matter model.} \cite{freese}. 
These observations are consistent with a mirror matter halo because the 
entire halo would not be expected to be in the form of mirror stars. 
Mirror gas and dust would also be expected because they are a necessary 
consequence of stellar evolution and should therefore significantly populate 
the halo. 
Thus, perhaps MACHOs are the first observational evidence of the mirror 
matter. 

It is also plausible that in the galactic halo some fraction of mirror stars 
exists in the form of compact substructures like globular or open clusters, in 
the same way as it happens for ordinary stars. 
In this case, for a significant statistics, one could observe interesting time 
and angular correlations between the microlensing events.

%---------------------------------------------------------------------------------

\section{The stellar models}
\label{mirror_star_mod}

As we know, in the exact mirror symmetric scenario the microphysics of the 
hidden sector is exactly the same as the visible one, the only changes are 
due to the boundary conditions. 
This is a very favourable condition for the study of M stars, because the 
necessary knowledge is the same than for the O ones, that we know very 
well. 
This means that M stars follow the same evolutionary stages than visible 
ones. 
A very brief review of stellar evolution will be given at the beginning of the 
next section.

The same physics for both sectors means that the equations 
governing the mirror stellar evolution and the physical ingredients to put 
inside them (namely the equation of state, the opacity tables, and the 
nuclear reactions) are the same as for visible stars. 
The only change regards the composition of the M star. 
In fact, while the typical helium abundance for O stars is $ Y \simeq 0.24$, 
for the M stars we have $ Y' = 0.40$-$0.80 $. 
This interval is obtained considering that its lower limit is given by the 
primordial helium abundance coming from the mirror Big Bang 
nucleosynthesis studied in ref.~\cite{bcv}. 
In next section we will evaluate its impact on the evolution of M stars. 

If we consider a single isolated star \footnote{We are practically neglecting, 
as usual, the interactions existing in systems of two or three stars.}, 
its evolutionary and structural properties depend only on the mass and the 
chemical composition. 
In particular, the latter is expressed by the abundances by mass of hydrogen 
($ X $), helium ($ Y $), and the so-called heavy elements or metals ($ Z $), 
i.e. all the elements heavier than H and He, so that the condition $X+Y+Z=1$ 
is fulfilled.\footnote{From now on we will use the 
prime ($'$) to indicate mirror quantities only if they appear together with 
ordinary ones; otherwise we don't use it, taking in mind that high 
$ Y $-values refer to mirror stars and low $ Y $-values to the ordinary ones.}

We computed mirror star models using the FRANEC 
({\sl Frascati RAphson Newton Evolutionary Code}) evolutionary program, a 
numerical tool that solves the equations of stellar structures \cite{cassal97}. 
As physical inputs for this code we chose the opacity tables of 
ref.~\cite{alexfer94} for temperatures lower than 10000 $ K $ 
and those obtained in the Livermore laboratories and described in 
ref.~\cite{rogigl96} for higher temperatures, the equation of state presented in  
ref.~\cite{stran88}, implemented in the low-temperature regime 
with a Saha equation of state, and the Eddington approximation to the
grey atmosphere solution for the integration 
of stellar atmospheres\footnote{
The Eddington approximation assumes local thermodynamical 
equilibrium with opacity independent of frequency. 
In this case the temperature in the stellar atmosphere is given by
\vspace{-.2cm}
\begin{eqnarray}
T^4 (\tau)= {3 \over 4} T_{eff}^4 \left( \tau + {2 \over 3} \right) \;, \nonumber
\end{eqnarray}
where $ T(\tau) $ is the temperature of an atmospheric layer located at the 
optical depth $ \tau $, and $ T_{eff} $ is the effective temperature of the star.}. 
These inputs are valid over the entire ranges of temperatures and densities 
reached by our models.

For what concerns their evolutionary and structural properties, mirror 
stars are equivalent to ordinary stars with a very high helium abundance. 
Thus, we computed stellar models for large ranges of masses and helium 
contents, and for a low metallicity $ Z $. 

As far as it concerns the adopted value for this metallicity, we could adopt a 
value equal to zero as for \lq{normal}\rq\ Pop. III stars. 
However, also in the O sector we still lack of any empirical evidence for 
low-mass Pop. III stars, i.e. of stars with metallicity equal to zero. 
This result has been often explained as a consequence of the peculiar initial 
mass function\footnote{
The initial mass function gives the number of stars with mass in the range 
$M \div (M+dM)$ formed within a given stellar environment.} 
of Pop. III stars: the lack of metals should make less efficient the cooling 
processes within the primordial clouds so their fragmentation could produce 
only high-mass stars (whose evolutionary lifetimes are so short that they 
are no more observable. 
In the M sector, due to strong reduction of the H abundance, due to the huge 
increase of the He abundance, the cooling processes inside the primordial 
clouds should have also a lower efficiency, so only very massive M stars 
should form. 
For such reason, being interested to stellar objects that presently can 
\lq{work}\rq\ as MACHOs, we consider for present computation a metallicity 
$Z=10^{-4}$, very low but larger than zero, characteristic of the so-called 
stellar population II, i.e.~an old stellar population, coming soon after the 
first one.

%---------------------------------------------------------------------------------

\section{Evolution of the mirror stars}
\label{mirror_star_evol}

First of all we remember that in stellar astrophysics the evolution of a star is 
studied in the so-called {\sl H-R (Hertzsprung-Russell) diagram}, where we 
plot the luminosity $ L $ and the effective temperature $ T_{eff} $ of the 
star\footnote{The {\sl effective temperature} $ T_{eff} $ of a star is defined 
by 
\begin{eqnarray}
L = 4 \pi R^2 \sigma T_{eff}^4 \;, \nonumber
\end{eqnarray}
where $ \sigma $ is the Stephan-Boltzmann constant, $ L $ is the 
luminosity, and $ R $ is the radius at the height of the photosphere. 
Thus, $ T_{eff} $ is the characteristic temperature of the stellar surface if it 
emits as a black body.}. 
In order to understand the evolutionary differences between ordinary and 
mirror stars, it is also necessary to give a very brief review of basic stellar 
evolutionary theory.

%  FIGURA  %
\begin{figure}[p]
  \begin{center}
    \leavevmode
    \epsfxsize = 12cm
    \epsffile{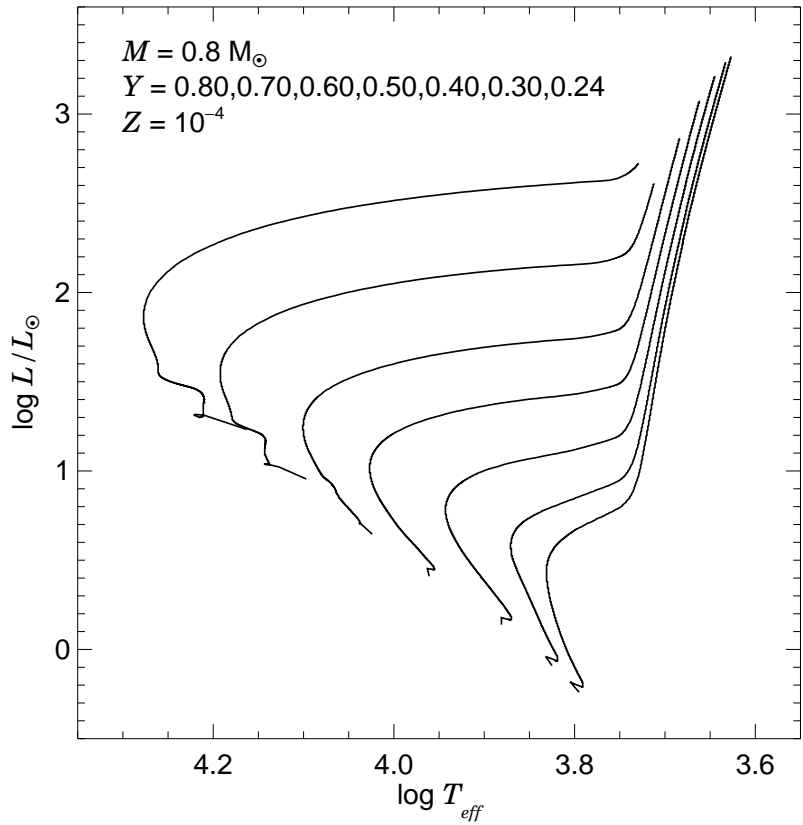}
  \end{center}
\vspace{-2.cm}
\caption{\small Evolutionary tracks in the H-R diagram of stars with 
$ M = 0.8M_\odot $, $ Z = 10^{-4} $ and different helium contents 
$ Y = 0.24, 0.30, 0.40, 0.50, 0.60, 0.70, 0.80 $. 
The last (brightest) point along each evolutionary track corresponds to the 
He ignition stage (see text for more details).}
\label{mirstar_fig_tesi_1}
\end{figure}
% --------------- %

During the first fast phase of gravitational contraction at nearly constant 
effective temperature and decreasing luminosity, the star goes down along 
its Hayashi track\footnote{
{\sl Hayashi track} is the evolutionary track of a 
totally convective stellar model. 
It is the coldest possible track for a star of a given mass, and it is located at 
the extreme right of the H-R diagram.}, 
negligibly slowing down its 
contraction only while the structure is fast burning the few light elements 
(D, Li, Be, B) present. 
Contraction increases the central temperature $ T_c $, until stars with 
masses $ M \gsim 0.1 ~M_\odot $ \footnote{The exact value of the 
{\sl hydrogen burning minimum mass} $ M_{hbmm} $ is dependent on the 
metallicity. 
For our models $ Z = 10^{-4} $ and $ M_{hbmm} \simeq 0.1 ~M_\odot $, while 
for solar metallicity $ Z = 0.02 $ and $ M_{hbmm} \simeq 0.08 ~M_\odot $.} 
ignite the hydrogen burning\footnote{There are two possible ways of burning 
hydrogen: the first one, called PP or proton-proton chain, becomes efficient 
at $ T_c \sim 6 \times 10^6 ~{\rm K} $, while the second one, called CNO 
chain, at $ T_c \sim 15 \times 10^6 ~{\rm K} $. 
Since their efficiencies are dominant at different temperatures, the PP chain 
provides energy for smaller stellar masses, and the CNO does it for bigger 
ones.} in their cores at a temperature $ T_c \sim 6 \times 10^6 ~{\rm K} $, 
while the ones with lower masses do not ignite hydrogen and die as 
{\sl brown dwarfs}. 
After depletion of hydrogen in the core, the burning passes to a shell at the 
boundary of the core, which is now made of He. 
At this stage the star starts to decrease its effective temperature 
({\sl turn-off}). 
Meanwhile the He core contracts until stars with $ M \gsim 0.5 ~M_\odot $ 
reach a central temperature $ T_c \sim 10^8 ~{\rm K} $ 
and start He-burning into C and O; stars with lower masses die as {\sl white 
dwarfs} and start their {\sl cooling sequences}. 
The stars with mass larger than 6-8$ M_\odot $ are able to ignite the 
successive nuclear burning processes and die exploding as type II 
{\sl supernovae} leaving in their place a {\sl neutron star} or a {\sl black 
hole}. 

\begin{table}[p]
\begin{center}
\begin{tabular}{|c||c|c|c|c|c|c|c|c|} \hline \hline
  $ Y $ & 0.24 & 0.30 & 0.40 & 0.50 & 0.60 & 0.70 & 0.80 \\
  \hline
  age ($ 10^9 yr $) & 12.4 & 8.53 & 4.50 & 2.17 & 1.01 & 0.417 & 0.169 \\
  \hline \hline
\end{tabular}
\end{center}
\caption{\small Ages computed for stars of mass $ M = 0.8 M_\odot $ and 
the indicated helium contents.}
\label{table2_7}
\end{table}

%  FIGURA  %
\begin{figure}[p]
  \begin{center}
    \leavevmode
    \epsfxsize = 10cm
    \epsffile{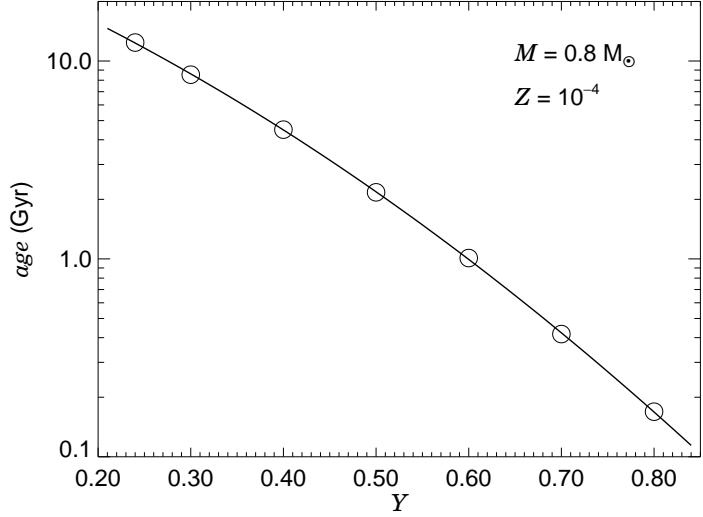}
  \end{center}
\vspace{-5.cm}
\caption{\small Evolutionary times (listed in table \ref{table2_7}) for models 
with $ M = 0.8 \; M_\odot $, $ Z = 10^{-4} $ and different helium contents 
$ Y = 0.24, 0.30, 0.40, 0.50, 0.60, 0.70, 0.80 $.
The line shows the quadratic fit.}
\label{mirstar_fig_tesi_3}
\end{figure}
% --------------- %

A key point is the evaluation of evolutionary times. 
For any given stellar mass, the evolutionary phase with the longest lifetime 
is the one corresponding to the central hydrogen burning stage 
(the so-called {\sl main sequence}) with 
time-scales of $ 10^{10} $ yr for masses near the solar mass. 
Thus, we can approximate the lifetime of a star with its main sequence time. 
Since both luminosity and effective temperature depend on the mass and 
chemical composition, clearly the lifetime too depends on them. 
We use now the proportionality relations valid for low mass stars \cite{claybook}
\begin{equation}
L \propto \mu^{7.5} M^{5.5}
\label{lmum}
\end{equation}
and
\begin{equation}
T_{eff}^4 \propto \mu^{7.5} \;,
\label{teffmu}
\end{equation}
where $ \mu $ is the mean molecular weight. 
From eq.~(\ref{lmum}) we obtain that bigger masses need higher 
luminosities, so that they use all the available hydrogen earlier than the 
lighter ones. 
From both eqs.~(\ref{lmum}) and (\ref{teffmu}) we know that an increase of 
helium abundance corresponds to an increase of the mean molecular weight 
and consequently in both luminosity and effective temperature. 
The increase in luminosity means that the star needs more fuel to produce 
it, but at the same time its amount is lower, because higher $ Y $ values 
necessarily imply lower $ X $ values. 
Both these events act to shorten the lifetime of a mirror star, which has a 
high He content. 
This can be formalized in the following relation \cite{claybook}
\begin{equation}
{\rm t_{MS}} \propto { X \over \mu^{1.4} } 
                    \sim X \left({5X+3 \over 4} \right)^{1.4} \;,
\label{txmu}
\end{equation}
where $ t_{MS} $ is the main sequence lifetime.

%  FIGURA  %
\begin{figure}[p]
  \begin{center}
    \leavevmode
    \epsfxsize = 12cm
    \epsffile{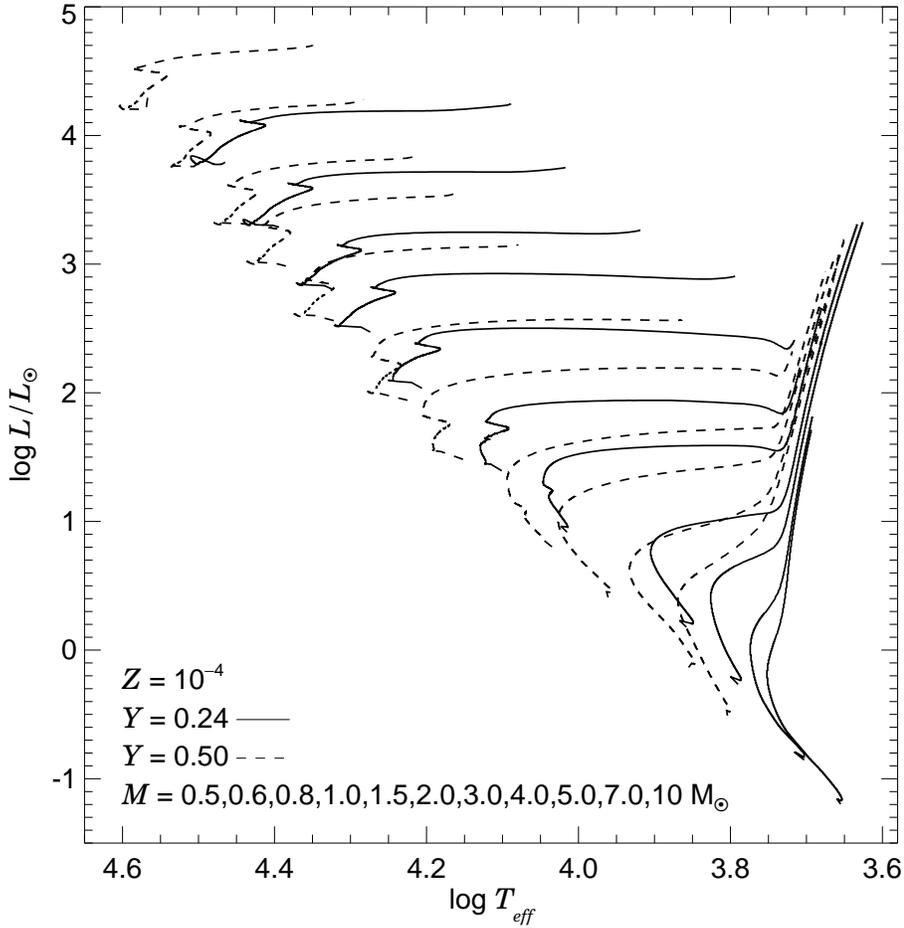}
  \end{center}
\vspace{-2.cm}
\caption{\small Evolutionary tracks in the H-R diagram of stars with different 
masses $ M = 0.5, 0.6, 0.8, 1.0, 1.5, 2.0, 3.0, 4.0, 5.0, 7.0, 10 \; M_\odot $, 
$ Z = 10^{-4} $ and two different helium contents $ Y = $ 0.24 and 0.50.}
\label{mirstar_fig_tesi_2}
\end{figure}
% --------------- %

The average change of the effective temperature is a consequence of the 
fact that the radiative opacity of a He-rich mixture is lower, and a lower 
opacity produces hotter effective temperatures.

After these predictions on evolutionary properties of He-rich stars, we 
analyze the quantitative results of our models. 
They can be divided into two groups. 
The first one is made of models with mass $ M = 0.8 M_\odot $ and many 
different $ Y $ values. 
The second one is made instead of models with only three different He 
contents and a large range of masses.

We start from figure \ref{mirstar_fig_tesi_1}, where we plot the models of 
$ 0.8 M_\odot $ in the H-R diagram. 
The models are followed until the He-burning ignition, i.e. along their main 
sequence, turn-off and red giant\footnote{A red giant is a cold giant star in 
the phase of H-burning in shell before the He-ignition.} phases, which 
practically occupy all their lifetimes. 
Our qualitative predictions are indeed confirmed. 
Models with more helium are more luminous and hot; for example the main 
sequence luminosity ratio of the model with $ Y = 0.80 $ to the one with 
$ Y = 0.24 $ is $ \sim 10^2 $. 
Other consequences of an He increase are a longer (in the diagram, not in 
time) phase of decreasing temperature at nearly constant luminosity, and a 
shorter red giant branch. 

From these models we computed the evolutionary times until the 
He-ignition, i.e. for the entire plotted tracks, and we summarize them in 
table \ref{table2_7}. 
As expected, the ages decrease for growing $ Y $, but we see now how much 
high is this correlation. 
For $ Y = 0.40 $ the lifetime is already about one third compared to a visible 
($ Y = 0.24 $) star, while for the highest value, $ Y = 0.80 $, it is roughly 
$ 10^2 $ times lower. 
We can approximately say that an increase of $ 10 \% $ in helium 
abundance roughly divides by two the stellar lifetime. 
In figure \ref{mirstar_fig_tesi_3} we plot the evolutionary times listed in the 
table. 
We see that, using a logarithmic scale for the stellar age, we obtain a 
quasi-linear relation between it and the helium content for this range of 
parameters. 
An almost perfect approximation (the standard deviation for $\log(age)$ is 
$\sigma_{\rm age} = 0.0055 $) can be obtained with a quadratic fit, which 
gives the following expression for the age expressed in Gyr

\begin{equation}
{\rm log}\,(age) = 1.61 -1.81\: Y - 1.46\: Y^2
\;\;\;\;\;\;\;\;\;\;\;\; {\rm for} \;\; M = 0.8 \: M_{\rm \odot} \;\; .
\label{logt_y}
\end{equation}

\begin{table}[p]
\begin{center}
\begin{tabular}{|c|c|c|c|} \hline \hline
  mass & age ($ 10^9 yr $) & age ($ 10^9 yr $) & age ($ 10^9 yr $) \\
  ($ M/M_\odot $) & ($ Y = 0.24 $) & ($ Y = 0.50 $) & ($ Y = 0.70 $) \\
  \hline \hline
  0.5 & $  66.7  $ & $  11.2  $ & $  1.92  $ \\  
\hline
  0.6 & $  35.0  $ & $  5.80  $ & $  1.04  $ \\  
\hline
  0.8 & $  12.4  $ & $  2.17  $ & $  0.417  $ \\  
\hline
  1.0 & $  5.73  $ & $  1.05  $ & $  0.219  $ \\  
\hline
  1.5 & $ 1.50  $ & $  0.301  $ & $ 0.0902  $ \\  
\hline
  2.0 & $  0.608  $ & $  0.140  $ & $  0.0445  $ \\  
\hline
  3.0 & $  0.204  $ & $  0.0564  $ & $  0.0178  $ \\  
\hline
  4.0 & $  0.110  $ & $  0.0313  $ & $  0.00941  $ \\  
\hline
  5.0 & $  0.0697  $ & $  0.0205  $ & $  0.00656  $ \\  
\hline
  7.0 & $  0.0366  $ & $  0.0117  $ & $  0.00414  $ \\  
\hline
  10 & $  0.0202  $ & $  0.00718  $ & $  0.00278  $ \\
  \hline \hline
\end{tabular}
\end{center}
\caption{\small Ages computed for stars of the indicated masses and helium 
contents, with a metallicity $ Z = 10^{-4} $.}
\label{table3_7}
\end{table}

%  FIGURA  %
\begin{figure}[p]
  \begin{center}
    \leavevmode
    \epsfxsize = 10cm
    \epsffile{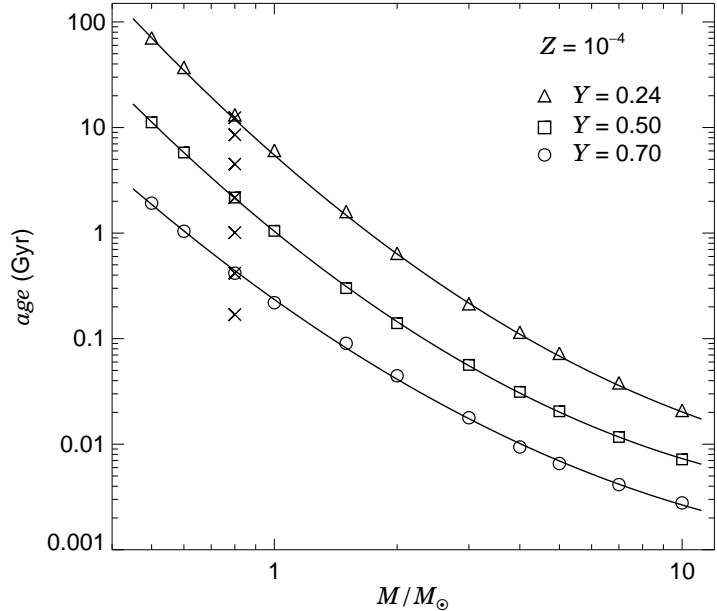}
  \end{center}
\vspace{-4.cm}
\caption{\small Evolutionary times (listed in table \ref{table3_7}) for models 
with $ M = $ 0.5, 0.6, 0.8, 1.0, 1.5, 2.0, 3.0, 4.0, 5.0, 7.0, 10$ \; M_\odot $, 
$ Z = 10^{-4} $ and three different helium contents $ Y = $ 0.24, 0.50 and 
0.70. 
Are also indicated the models of figure \ref{mirstar_fig_tesi_3} and the fitted 
curves.} 
\label{mirstar_fig_tesi_4}
\end{figure}

Let us now extend the analysis to models covering a large range of masses, 
from 0.5 $ M_\odot $ to 10 $ M_\odot $. 
Since the dependence on the helium content has been already studied for 
the 0.8 $ M_\odot $ case, we concentrate on only three $ Y $ values. 
Figure \ref{mirstar_fig_tesi_2} shows the evolutionary tracks for all the 
masses and only two helium contents, and again the models are followed up 
to the He-ignition. 
For every mass the $ Y $ dependence is the same as for the above 
discussed 0.8 $ M_\odot $ model. 
For models with masses $ M \gsim 2 ~M_\odot $ the growth in $ Y $ causes 
a considerable increase of the He-ignition effective temperature, together 
with the disappearance of the red giant branch.

In table \ref{table3_7} we list the lifetimes for all masses and the three 
indicated helium contents. 
The ratio of an ordinary star ($ Y = 0.24 $) evolutionary time to the high 
He-content mirror one ($ Y = 0.70 $) is between $ \sim 30 $ for 0.5 
$ M_\odot $ and $ \sim 10 $ for 10 $ M_\odot $. 
These data are plotted in figure \ref{mirstar_fig_tesi_4}, where we see that 
the same dependence on the star masses holds for every helium content, 
with a shift toward lower ages for higher $ Y $ values.
In this case we perform a quadratic fit using logarithmic scales in both 
coordinates and with age expressed in Gyr, obtaining 

\begin{equation}
{\rm log}\,(age) = 0.738 - 3.398\: {\rm log}\,M + 0.966\: ({\rm log}\,M)^2 
\;\;\;\;\;\;\;\;\;\;\;\; {\rm for} \;\; Y=0.24 \;\; ,
\label{logt_logm1}
\end{equation}
\begin{equation}
{\rm log}\,(age) = 0.016 - 3.143\: {\rm log}\,M + 0.986\: ({\rm log}\,M)^2 
\;\;\;\;\;\;\;\;\;\;\;\; {\rm for} \;\; Y=0.50 \;\; ,
\label{logt_logm2}
\end{equation}
\begin{equation}
{\rm log}\,(age) = -0.630 - 2.750\: {\rm log}\,M + 0.802\: ({\rm log}\,M)^2 
\;\;\;\;\;\;\;\;\;\, {\rm for} \;\; Y=0.70 \;\; ,
\label{logt_logm3}
\end{equation}
where the standard deviations $\sigma_{\rm age}$ for $\log(age)$ are 
respectively 0.018, 0.010, 0.029.
We can also consider together the dependencies of stellar ages on both 
the helium content and stellar masses, obtaining 
\begin{eqnarray}
{\rm log}\,(age) = 1.31 - 2.16\: Y - 3.73\: {\rm log}\,M - 0.86\: Y^2 + \nonumber \\
  + 1.34\: Y {\rm log}\,M + 0.99\: ({\rm log}\,M)^2 - 0.28\: Y^2 ({\rm log}\,M)^2 \;\;.
\label{logt_y_logm}
\end{eqnarray}

%  FIGURA  %
\begin{figure}[p]
  \begin{center}
    \leavevmode
    \epsfxsize = 12cm
    \epsffile{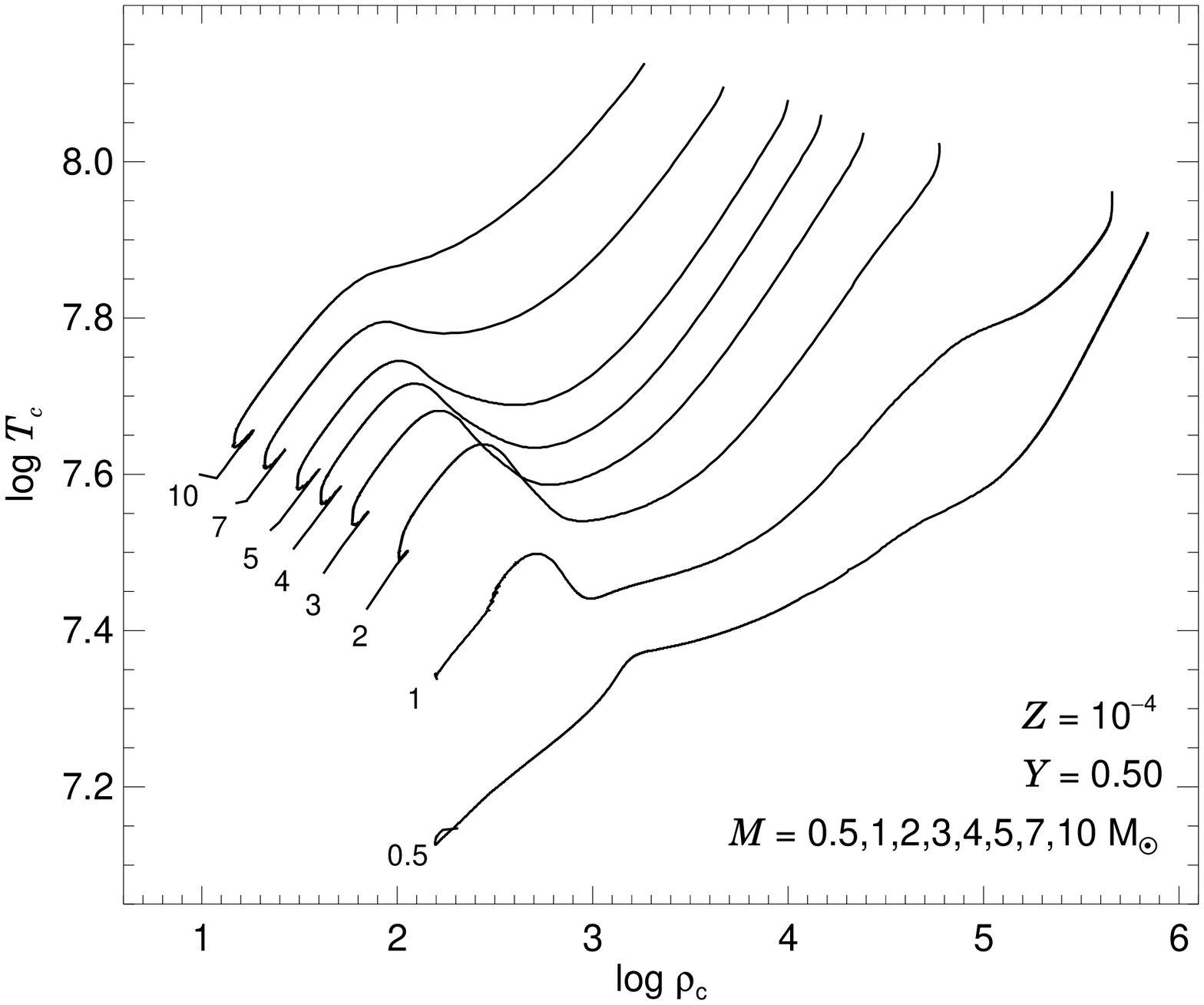}
  \end{center}
\vspace{-3.5cm}
\caption{\small Behaviour of the central temperature $T_c$ as a function of 
the central density $ \rho_c $ during the evolution for stars with different 
masses: $ M = 0.5, 1.0, 2.0, 3.0, 4.0, 5.0, 7.0, 10 \; M_\odot $, 
$ Z = 10^{-4} $, and $ Y = $ 0.50.}
\label{evol_fig_12}
\end{figure}

%  FIGURA  %
\begin{figure}[p]
  \begin{center}
    \leavevmode
    \epsfxsize = 12cm
    \epsffile{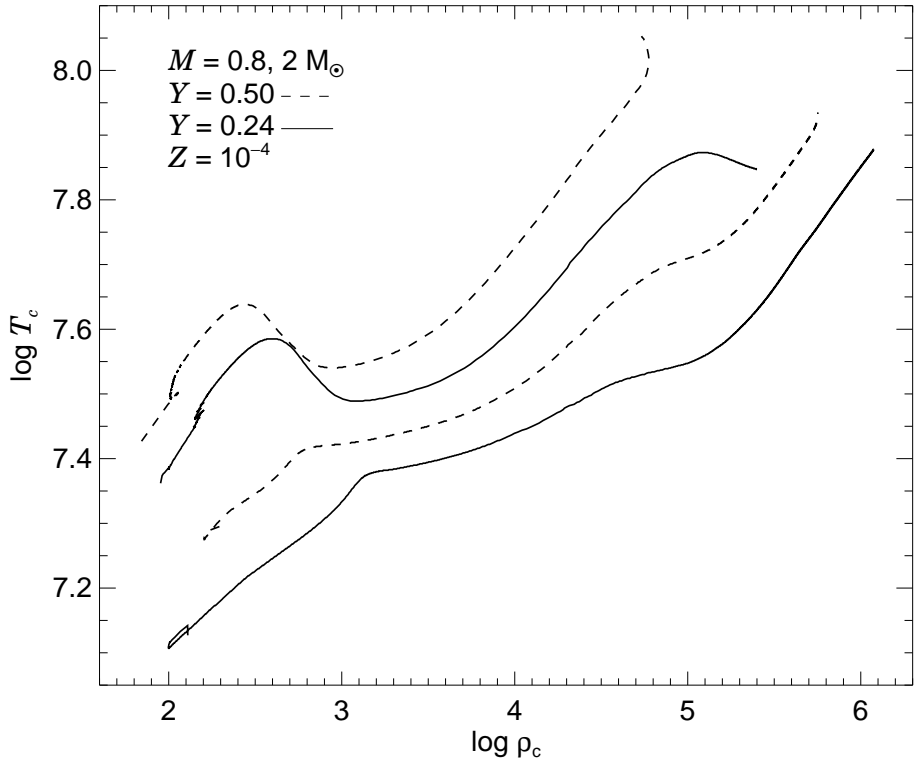}
  \end{center}
\vspace{-3.5cm}
\caption{\small The trend of the central temperatures $ T_c $ as a function of 
the central density $ \rho_c $ of two stellar models with masses equal to 
$0.8M_\odot$ and $2M_\odot$, metallicity $ Z = 10^{-4} $ and initial He
contents equal to $ Y = 0.24 $ and 0.50.}
\label{evol_fig_13}
\end{figure}
% --------------- %

%  FIGURA  %
\begin{figure}[p]
  \begin{center}
    \leavevmode
    \epsfxsize = 12cm
    \epsffile{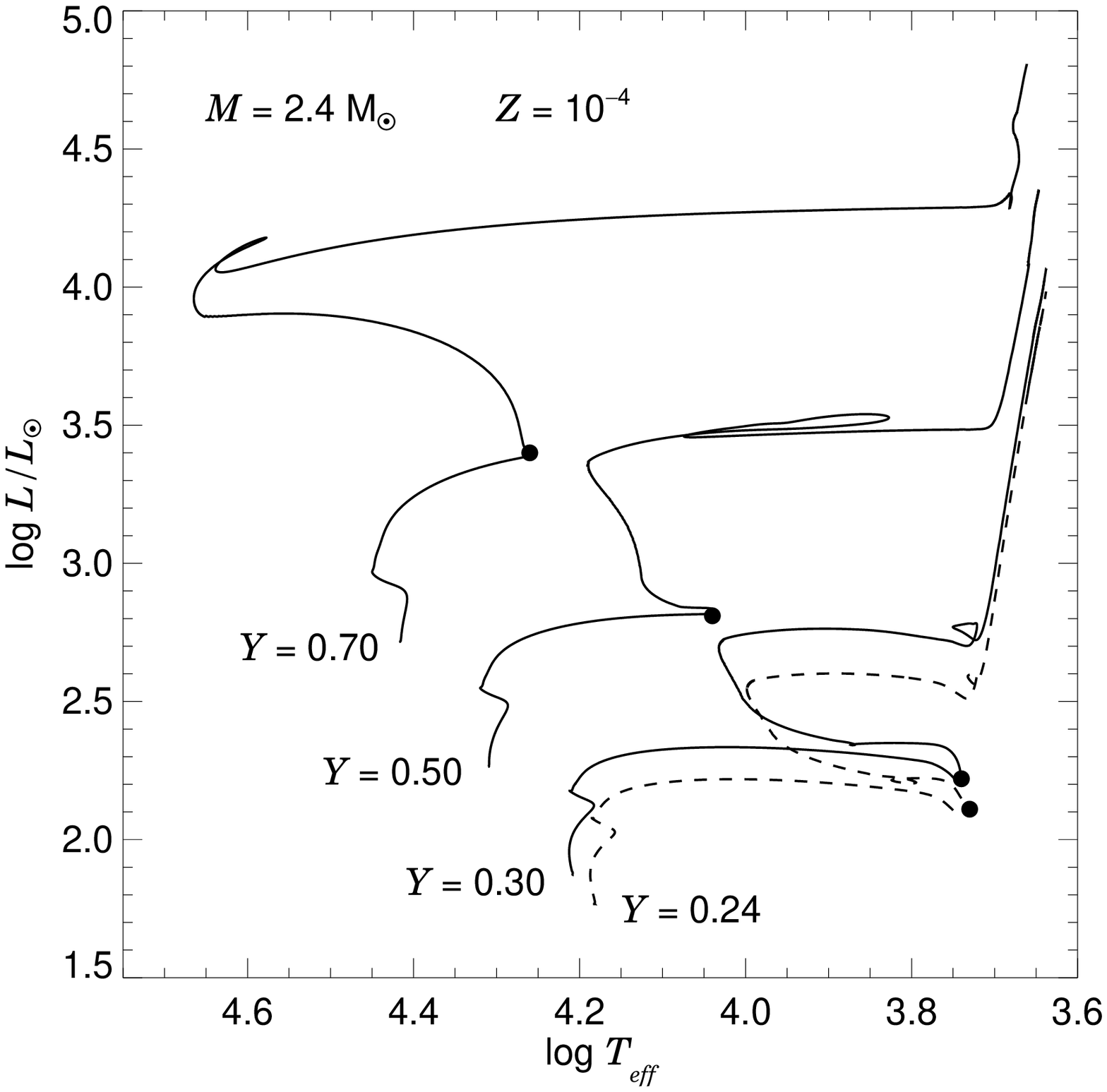}
  \end{center}
\vspace{-2.cm}
\caption{\small Evolutionary tracks in the H-R diagram of stars with 
$ M = 2.4 M_\odot $, $ Z = 10^{-4} $ and different helium contents 
$ Y = 0.24, 0.30, 0.50, 0.70 $. 
The filled circles along each evolutionary track correspond to the 
He ignition stage (see text for more details).}
\label{evol_fig_15}
\end{figure}
% --------------- %

This is an evidence that, under large mass ranges and different boundary 
conditions (in terms of temperatures of the mirror sector, and thus stellar 
helium contents), the lifetimes of mirror stars are roughly an order of 
magnitude smaller than the ones of visible stars. 
This means that, compared to O stars, M stars evolve faster and enrich 
earlier the galaxy of processed mirror gas, with implications for galaxy 
evolution. 

In fig.~\ref{evol_fig_12}, we show the behaviour of the central temperature as 
a function of the central density during the main H-burning phase for the M 
star models with an He abundance equal to Y=0.50. 
In order to allow to evaluate the effects of the initial He abundance on the 
behaviour with time of the thermal conditions at the center of the star, in 
fig.~\ref{evol_fig_13} we plot the behaviour of the central temperature as a 
function of the central density for two selected stellar masses, 0.8 and 2 
$M_\odot$, for different assumptions about the initial He content.

It is worth noticing that a larger He content has a huge effect on the central 
thermal conditions: larger the He content, hotter the star for any fixed value 
of the central density. 
The effect is larger for less massive stars, i.e. $M \le 1.0 M_\odot$, with 
respect the more massive. 
This evidence has to be related to the different H-burning mechanism
at work in the different ranges of mass: in low-mass stars H-burning occurs 
mostly through the {\sl p-p} chain which has a lower temperature 
dependence with respect the {\sl CNO} cycle (occurring in more massive 
stars). 
So, when the initial He content increases, in order to fulfil the stellar energy 
requirement, the central temperature has to increase larger than in more 
massive stars.

One has also to note that the significant increase of the central temperature, 
when Y increases, has the quite important consequence that for initial He 
abundance of the order of $\sim0.40$ also 
low-mass stars with mass around $0.8M_\odot$ reaches in their interiors 
the temperature required to activate the {\sl CNO} cycle 
($T\approx14\times10^6~K$). 
When this happens the H-burning occurs in a convective core. 
This occurrence has a further consequence which will be discussed in the 
following.  

The huge increase of the central temperature with the initial He content has a 
strong impact also on the evolutionary phases successive to the core 
H-burning one: as the central temperature increases the thermal conditions 
requested in order to ignite He-burning are achieved sooner after the central 
H-exhaustion. 
This means that intermediate-mass stars, i.e. those structures 
which do not develop conditions of electron degeneracy inside their He core, 
are able to start burning He at quite larger effective temperature as already 
shown (see also fig.~\ref{evol_fig_15}). 

In low-mass stars, the physical behaviour is also more complicated. 
In the O sector, these structures, at the end of the core H-burning phase 
develop conditions of strong electron degeneracy inside their He core. 
As a consequence they are forced to largely increase the mass of the He 
core, through the H-burning occurring in a shell surrounding the He core, until 
the energy released by the gravitational contraction does overcome the 
energy losses due to thermal conduction and plasma neutrinos. 
This occurs near the tip of the Red Giant Branch (RGB), where the thermal 
conditions for He ignition are achieved and the star starts to burn helium 
through a violent He flash\footnote{
The He ignition is a violent process since it occurs inside a He core that is 
under conditions of strong electron degeneracy.}.

In the M sector, the large initial He content has the consequence that the 
structures are always hotter than O stars.
So, all along their evolution, low-mass  M stars develop electron 
degeneracy at a quite lower level. 
This occurrence has the important effect of strongly reducing the mass of 
the He core at the tip of the RGB and, in turn, the brightness of the tip
(so the extension in luminosity of the RGB is significantly reduced\footnote{
The brightness of the point corresponding to the He ignition is strongly 
dependent on the mass of the He core. 
This occurrence provides a direct explanation of the behaviour of the 
brightest point along the tracks plotted in fig.~\ref{mirstar_fig_tesi_1}, with 
the initial He content.}).

\begin{figure}[p]
  \begin{center}
    \leavevmode
    \epsfxsize = 10cm
    \epsffile{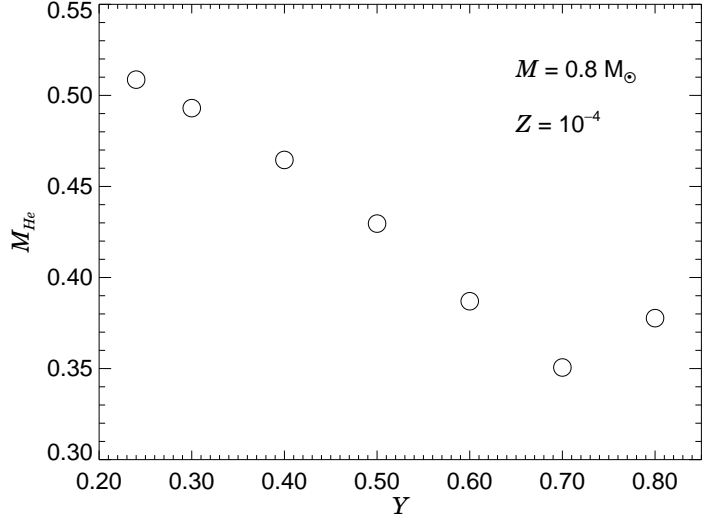}
  \end{center}
\vspace{-5cm}
\caption{\small He-core masses (listed in table \ref{table_mhe}) computed 
at He ignition for models with $ M = 0.8 \; M_\odot $, $ Z = 10^{-4} $ and 
different helium contents $ Y = 0.24, 0.30, 0.40, 0.50, 0.60, 0.70, 0.80 $.}
\label{evol_fig_9}
\end{figure}

In fig.~\ref{evol_fig_9} and table \ref{table_mhe}, we show the trend 
of the mass of the He core at the 
RGB tip, i.e. at the He ignition, as a function of the initial He abundance: 
the reduction of the He core mass when increasing the He content from 
$\sim0.2$ to $\sim0.7$ 
is really quite large $\sim 0.15M_\odot$. 
However, when increasing the value of Y from 0.70 to 0.80, we notice that 
the He core mass does increase at odds with previous indications. 
This is the consequence of the fact that increasing Y to quite large values, 
causes that the H-burning occurs via {\sl CNO} cycle so in a convective core. 
In addition, the stellar interiors are so hot that they are no more affected by 
electron degeneracy. 
The $0.8M_\odot$ model computed by assuming $Y=0.80$ is so hot that it 
burns H in a convective core and does not develop electron degeneracy at 
the central H-exhaustion, so its mass of the He core at the He ignition is no 
more fixed by the level of electron degeneracy but it depends on the 
maximum extension of convective core during the core H-burning phase.

\begin{table}[p]
\begin{center}
\begin{tabular}{|c||c|c|c|c|c|c|c|c|} \hline \hline
  $ Y $ & 0.24 & 0.30 & 0.40 & 0.50 & 0.60 & 0.70 & 0.80 \\
  \hline
  $M_{He} (M_{\odot})$ & 0.509 & 0.493 & 0.465 & 0.430 & 0.387 
  & 0.351 & 0.378 \\
  \hline \hline
\end{tabular}
\end{center}
\caption{\small He-core masses computed at He ignition for stars 
of mass $ M = 0.8 M_\odot $ and the indicated helium contents.}
\label{table_mhe}
\end{table}

%  FIGURA  %
% --------------- %

\begin{figure}[p]
  \begin{center}
    \leavevmode
    \epsfxsize = 10cm
    \epsffile{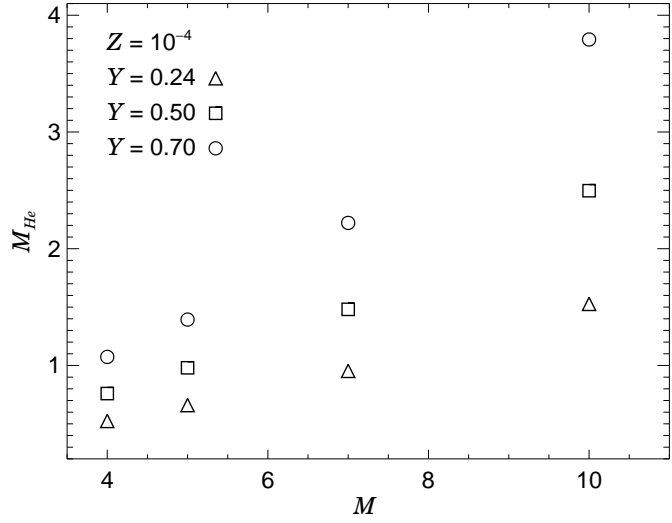}
  \end{center}
\vspace{-5cm}
\caption{\small He-core masses computed 
at He ignition for models (listed in table \ref{table_mhe2}) for models 
with $ M = $ 4.0, 5.0, 7.0, 10$ \; M_\odot $, $ Z = 10^{-4} $ and three different 
helium contents $ Y = $ 0.24, 0.50 and 0.70. }
\label{evol_fig_10}
\end{figure}

In more massive stars which also in the O sector do not develop conditions 
of electron degeneracy in their He core, this occurrence is also more 
evident (see table~\ref{table_mhe2} and fig.~\ref{evol_fig_10}). 
In fact, increasing the initial He content, the mass size of the convective 
core during the central H-burning stage increases as a consequence of the 
larger radiative flux (we remember that He-rich stars are brighter and then 
more energy has to be produced via nuclear burning in the interiors), and in 
turn the mass of the He core at the He ignition largely increases.

\begin{table}[p]
\begin{center}
\begin{tabular}{|c|c|c|c|} \hline \hline
  mass & $M_{He} (M_{\odot})$ & $M_{He} (M_{\odot})$ & 
$M_{He} (M_{\odot})$ \\
  ($ M/M_\odot $) & ($ Y = 0.24 $) & ($ Y = 0.50 $) & ($ Y = 0.70 $) \\
  \hline \hline
  4.0 & $ 0.524 $ & $ 0.760 $ & $ 1.07 $ \\  
\hline
  5.0 & $ 0.660 $ & $ 0.980 $ & $ 1.39 $ \\  
\hline
  7.0 & $ 0.953 $ & $ 1.48 $ & $ 2.22 $ \\  
\hline
  10 & $ 1.53 $ & $ 2.50 $ & $ 3.79 $ \\
  \hline \hline
\end{tabular}
\end{center}
\caption{\small He-core masses computed at He ignition for stars of the 
indicated mass and helium content, with a metallicity $ Z = 10^{-4} $.}
\label{table_mhe2}
\end{table}

%  FIGURA  %
% --------------- %

\begin{table}[p]
\begin{center}
\begin{tabular}{|c||c|c|c|c|c|c|c|c|} \hline \hline
  $ Y $ & 0.24 & 0.30 & 0.40 & 0.50 & 0.60 & 0.70 & 0.80 \\
  \hline
  $M^{up} (M_{\odot})$ & 6.3  & 5.8 & 4.8 & 3.7 & 2.6 & 1.8 & 1.6 \\
  \hline \hline
\end{tabular}
\end{center}
\caption{\small $M_{up}$ masses computed for models with $ Z = 10^{-4} $ 
and the indicated helium contents.}
\label{table_mup}
\end{table}

%  FIGURA  %
\begin{figure}[p]
  \begin{center}
    \leavevmode
    \epsfxsize = 10cm
    \epsffile{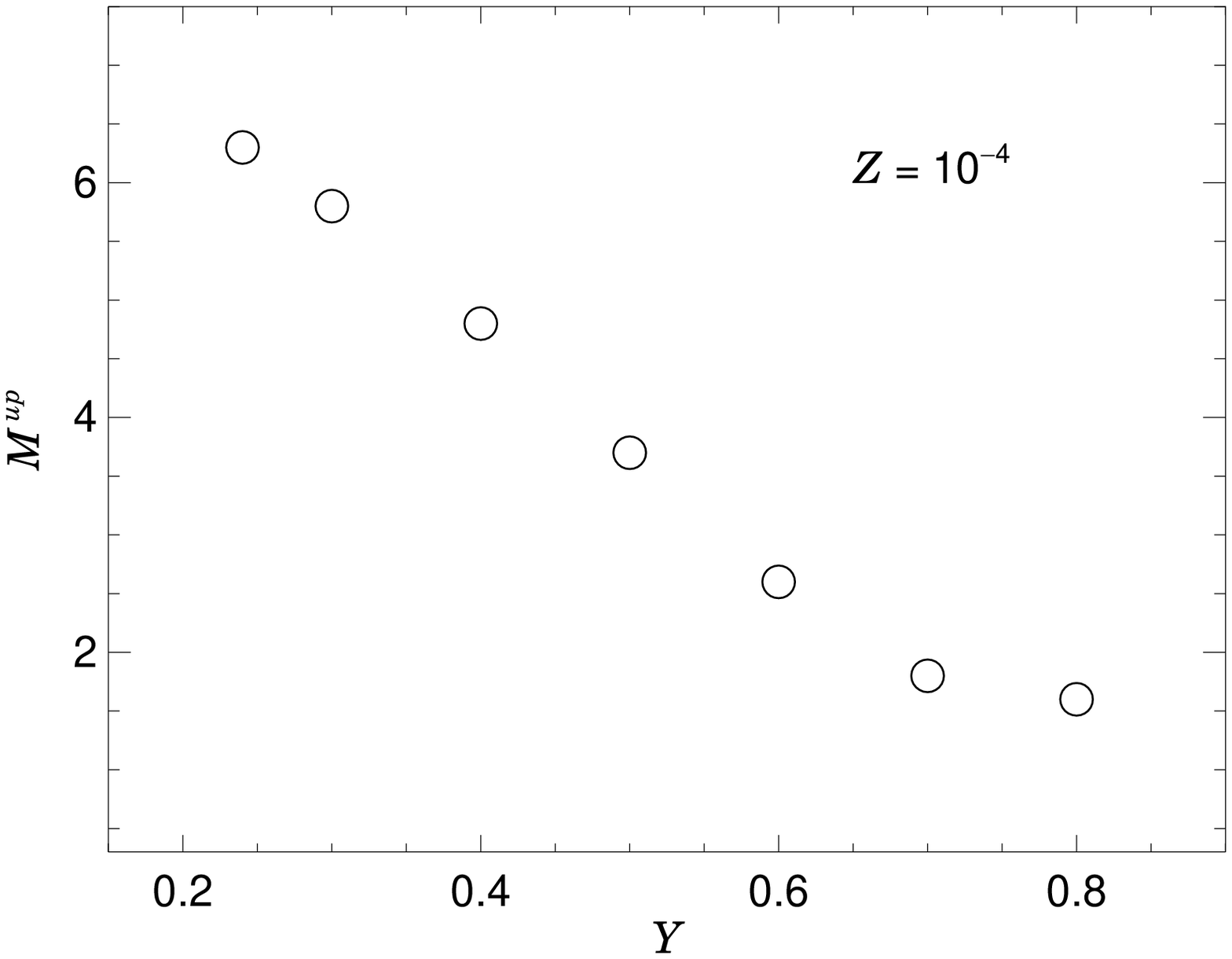}
  \end{center}
\addvspace{-5.cm}
\caption{\small $M^{up}$ masses (listed in table \ref{table_mup}) computed 
for models with $ Z = 10^{-4} $ and helium contents $ Y = $ 0.24, 0.30, 0.40, 
0.50, 0.60, 0.70 and 0.80. }
\label{evol_fig_11}
\end{figure}
% --------------- %

For any given stellar population, i.e. for each fixed initial chemical 
composition, one of the most important parameters characterizing that 
stellar population is the value of the critical mass ($M^{up}$) between the 
stars which after the core He-exhaustion are not able to ignite the 
successive nuclear burning processes so becoming carbon-oxygen white 
dwarfs, and those structures massive enough to ignite carbon and all 
successive nuclear burnings so concluding their life as type II supernovae.

Figure~\ref{evol_fig_11} and table \ref{table_mup} show the trend of 
$M^{up}$ as a function of the initial He content: 
it is worth noting the quite relevant reduction in the value of this critical 
mass when increasing the He content from $\sim0.20$ to $\sim0.8$. 
This occurrence is due to the convolution of two effects related to the 
increase of the He abundance: 1) the increase of the He core at the He 
ignition, 2) the larger temperatures of the stellar interiors.

The huge reduction of the value of $M^{up}$ has the important consequence 
that in the M sector - when assuming an initial mass function similar to that 
of the O sector - the expected fraction of stars exploding as type II 
supernovae should be quite larger than that expected in the O sector. 
In the meantime, the fraction of carbon-oxygen white dwarfs - which could be 
potentially contributors to the MACHOs population - would be significantly 
reduced.

From the detailed study of this evolution together with the necessary 
information of the initial mirror stellar mass function, one could predict the 
expected population of mirror stars, in order to compare it with current 
MACHO observations. 
In addition, one could also evaluate the amount of gravitational waves 
expected from supernovae in the mirror sector. 
These are just some examples of interesting future applications of the 
present study.

%---------------------------------------------------------------------------------

\section{Conclusions}

In this paper we have investigated the astrophysical consequences of the 
existence of a mirror sector on stellar scales.
In particular we studied in detail the evolutionary and structural properties 
of mirror stars, considered as very natural MACHO candidates.

Considering that BBN epoch in the mirror (M) world proceeds differently from 
the ordinary (O) one, and it predicts the mirror helium 
abundance in the range $Y' =0.4-0.8$, considerably larger than the 
observable $Y \simeq 0.24$, we studied the stars made of a large amount 
of helium (they can be both mirror stars or ordinary stars in a very late 
evolutionary phase of the Universe, when a lot of hydrogen already 
burned in helium). 

Using a numerical code, we computed evolutionary models of stars for high 
values of helium content and low metallicity ($ Z = 10^{-4} $), and for a large 
range of masses, spanning from 0.5 to 10 $ M_\odot $.

We found that, since stars in mirror sector have very different helium 
abundances from the visible ones (but the physics is the same), they have 
much faster evolutionary times, dependent on the exact He content and thus 
on the temperature of the mirror sector (since they are inversely 
proportional). 
The mean life of a mirror star can be until 30 times shorter than that of an 
ordinary one, if we consider the most helium rich stars.
Generally, we found the evidence that, under large mass ranges and  
different boundary conditions (in terms of temperatures of the mirror sector, 
and thus stellar helium contents), the lifetimes of mirror stars are roughly an 
order of magnitude smaller than the ones of visible stars. 

This means that, compared to the ordinary ones, mirror stars evolve faster 
and enrich earlier the galaxy of processed mirror gas, with implications for 
galaxy evolution. 
In addition, we found the important result that the minimum mass for carbon 
ignition, and then for the explosion of the star as type II supernova, is 
inversely proportional to the helium abundance, and hence for M stars it is 
lower than for the O ones; this means that in the M sector there are much 
more supernovae than in the visible one. 
This has important consequences on both the star formation rate 
induced by the shock waves coming from the supernova 
explosions, and the ratio of mirror star and gas in the halo of galaxies, with 
implications for the stability of a spherical halo.
In fact, the star formation reduces the mirror gas present, and the mirror halo 
looks like a collisionless system of stars, avoiding the M baryons to form 
disk galaxies as ordinary baryons do.
At the same time, the energy injected by the mirror supernovae explosions 
replace the one dissipated by the friction of the mirror gas and dust still 
present in the mirror galaxy, and helps to maintain its typical elliptical 
structure.
We notice that both faster evolutionary times and greater number of 
supernova events have the same effect of an accelerated chemical 
evolution on the mirror galaxy respect to the observable one.

From the evolutionary data obtained in the present work together with the 
necessary information on the initial mirror stellar mass function, we could 
predict the expected population of mirror stars, in order to compare it with 
current MACHO observations. 
In addition, we could evaluate the amount of gravitational waves expected 
from mirror supernovae, that are observable by the next detectors.

%---------------------------------------------------------------------------------

\section*{Acknowledgements}

\noindent 
The work is partially supported by the MIUR research grant on the Projects 
of National Interest PRIN2004 ``Astroparticle Physics''.

%---------------------------------------------------------------------------------

%---------------------------------------------------------------------------------

\end{document}